\documentclass[journal]{IEEEtran}
\usepackage{todonotes}
\usepackage[utf8]{inputenc}
\usepackage{amsmath,amssymb,amsfonts}
\usepackage{textgreek}
\usepackage{graphicx}
\usepackage{color,bm}
\usepackage{wasysym}
\usepackage{tikz}
\usepackage{cite}
\usepackage{array,multirow}
\usepackage{float}
\usepackage{subfigure}
\usepackage{balance}
\usepackage[hidelinks]{hyperref}
\usepackage{mathtools, nccmath}
\DeclareGraphicsExtensions{.tif}
\usepackage[linesnumbered,ruled,vlined]{algorithm2e}
\usepackage{amsmath} 
\usepackage{algorithmic,algorithm2e}

\newcount\Comments  % 1 suppresses notes to selves in text
\usepackage{color}
\definecolor{darkgreen}{rgb}{0,0.5,0}
\definecolor{purple}{rgb}{1,0,1}
\newcommand{\kibitz}[2]{\ifnum\Comments=1\textcolor{#1}{#2}\fi}
\newcommand{\first}[1]{\kibitz{black}{#1}}

\begin{document}

\title{Digital Twin-Assisted Cooperative Driving at Non-Signalized Intersections}

\author{Ziran~Wang,~\IEEEmembership{Member,~IEEE,}
        Kyungtae~Han,~\IEEEmembership{Senior Member,~IEEE,}
        and~Prashant~Tiwari,~\IEEEmembership{Member,~IEEE}% <-this % stops a space

\thanks{Z. Wang, K. Han, and P. Tiwari are with Toyota Motor North America R\&D, InfoTech Labs, 465 Bernardo Avenue, Mountain View, CA 94043 (e-mail: ryanwang11@hotmail.com; kyungtae.han@toyota.com; prashant.tiwari@toyota.com).}}% <-this % stops a space

\markboth{IEEE Transactions on Intelligent Vehicles, EARLY ACCESS}%
{Shell \MakeLowercase{\textit{et al.}}: Bare Demo of IEEEtran.cls for IEEE Journals}
\maketitle
\begin{abstract}
Digital Twin, as an emerging \first{technology related to Cyber-Physical Systems (CPS) and Internet of Things (IoT)}, has attracted increasing attentions during the past decade. \first{Conceptually, a Digital Twin is a digital replica of a physical entity in the real world, and this technology is leveraged in} this study to design a cooperative driving system at non-signalized intersections, \first{allowing} connected vehicles to cooperate with each other to cross intersections without any full stops. Within the proposed Digital Twin framework, we developed an enhanced first-in-first-out (FIFO) slot reservation algorithm to schedule the sequence of crossing vehicles, a consensus motion control algorithm to calculate vehicles' referenced longitudinal motion, and a model-based motion estimation algorithm to tackle communication delay and packet loss. Additionally, an augmented reality (AR) human-machine-interface (HMI) is designed to provide the guidance to drivers to cooperate with other connected vehicles. Agent-based modeling and simulation of the proposed system is conducted in Unity game engine based on a real-world map in San Francisco, and the human-in-the-loop (HITL) simulation results prove the benefits of the proposed algorithms with 20\% reduction in travel time and 23.7\% reduction in energy consumption, respectively, when compared with traditional signalized intersections.
\end{abstract}

% Note that keywords are not normally used for peer-review papers.

\IEEEpeerreviewmaketitle

\section{Introduction} \label{sec:Intro}
\subsection{Background}
\IEEEPARstart{B}{y} design, an intersection is a planned location where vehicles traveling from different directions may come into conflict, and its functional area extends upstream and downstream from the physical area of the crossing streets. A report from U.S. Department of Transportation indicates that, approximately 2.4 million intersection-related crashes occurred in 2007, representing 40\% of all reported crashes and 21.5\% of traffic fatalities \cite{fhwa2009intersection}. This shows that intersections represent a disproportionate share of the traffic safety issue. Among all intersection-related crashes, ``inadequate surveillance'' and ``false assumption of other's action'' are the top two driver attributed critical reasons, which are much lower ranked in non-intersection-related crashes \cite{nhtsa2010crash}.

Traffic signals have been playing a crucial role in achieving safer performance at intersections. Researchers and practitioners have shown that the appropriate installation and operation of traffic signals can reduce the severity of crashes \cite{fhwa2014signalized}. However, the addition of unnecessary or inappropriately-designed signals has adverse effects on traffic safety and mobility. In addition, the dual objectives of safety and mobility conflict in many cases. The design and operation of traffic signals at intersections require choosing elements that may lead to trade-offs in safety and mobility.

The emergence of connected vehicle technology during the past decades brings other possibilities to our transportation systems. Specifically, the level of connectivity within our vehicles has greatly increased, allowing these ``equipped'' vehicles to behave in a cooperative manner not only among themselves through vehicle-to-vehicle (V2V) communication, but also with other transportation entities through vehicle-to-infrastructure (V2I) communication, vehicle-to-network/cloud (V2N/V2C) communication, vehicle-to-pedestrian (V2P) communication, etc. (namely V2X communications). 

\first{The advancements of V2X communications makes the Digital Twin technology more feasible for connected vehicles. Although there is no universal definition of the Digital Twin, its basic concept is essentially the same across multiple studies \cite{glaessgen2012digital, lee2013recent}: A Digital Twin is a digital replica of a physical entity in the real world. With V2X communications, the Digital Twin technology enables real-time monitoring and synchronization of multiple real-world vehicles via their digital replicas. It has gained increasing momentum very recently in various connected vehicle studies \cite{alam2017c2ps, kumar2018novel, chen2018digital, wang2020adigital, liao2021cooperative}, and we also adopt this technology to develop a cooperative driving system at non-signalized intersections in this study. 
}

\subsection{Literature Review}
Cooperative driving at non-signalized intersections has been a popular topic in the research field of intelligent transportation systems along with the development of V2X communication. Since intersections are one of the most common traffic conflict situations, much work has been conducted to increase traffic safety and improve traffic flow by applying V2V communication and/or V2I communication. Most of the related work have been reviewed by Rios-Torres and Malikopoulos \cite{riostorres2017asurvey} and Wang et al. \cite{wang2020asurvey}, \first{where} some representative ones are covered in this subsection.

Pioneering work of cooperative driving at non-signalized intersections was conducted by Dresner et al., who proposed a multi-agent Autonomous Intersection Management (AIM) \cite{dresner2008amultiagent}. Their simulation study shows that their reservation-based approach outperforms current intersection designs (with traffic lights or stop signs). Fajardo et al. further developed a first-come-first-served (FCFS) reservation system for connected vehicles based on aforementioned study, where the microscopic traffic simulation shows the proposed system significantly reduces delay compared to a traditional traffic signal \cite{fajardo2011automated}. The reservation scheme have been also investigated by de La Fortelle \cite{fortelle2010analysis} and Zhang et al. \cite{zhang2015state}.

The ``virtual platoon'' concept was proposed by Neuendorf et al. with a decentralized motion controller designed for connected vehicles to cross the intersection \cite{neuendorf2004thevehicle}. Medina et al. further developed an automated intersection system with a similar approach, consisting of 1) a supervisory level with target vehicle assignment, and 2) an execution level with motion control design \cite{medina2018cooperative}. Some additional studies that adopt the similar virtual platoon methodology include the multi-agent motion management protocol developed by Jin et al. \cite{jin2013platoon}, and the distributed conflict-free cooperation proposed by Xu et al. \cite{xu2018distributed}.

Optimization-based approaches have also been widely adopted by many theoretical work in this research field. Related work can be basically categorized into several types, with the aims to optimize travel time \cite{li2006cooperative, yan2009autonomous, mirheli2019aconsensus}, minimize energy consumption \cite{malikopoulos2018adecentralized}, or a combination of both \cite{lee2012development, tlig2014decentralized, zhang2016optimal}. However, depending on how optimization problems are formulated, sometimes they could only be solved numerically with high computational load, which limits their real-time implementations \cite{riostorres2017asurvey}. 

In terms of real-world experimental implementations of the cooperative driving system at non-signalized intersections, Quinlan et al. designed a mixed reality platform of aforementioned AIM, where the vehicle can interact with multiple virtual vehicles in a simulation at a real intersection in real time \cite{quinlan2010bringing}. Additionally, a few experimental implementations were conducted by European research organizations, such as the Cybercars and Cybercars-2 projects in France \cite{bouraoui2006cybercar, fortelle2010analysis}. 

\subsection{Contributions of this Study}

Most existing theoretical and experimental research regarding the topic of cooperative driving at non-signalized (alternatively as ``unsignalized'' or ``signal-free'') intersections put their focuses on designing either the motion planning algorithms (to schedule the sequences of vehicles), or the motion control algorithm (to generate referenced motions of vehicles). Since they mostly assumed a pure connected and automated vehicle (CAV) environment (which will not be realistic in the near future), very few of them discussed the human-machine interface (HMI) design for drivers, namely how to provide the guidance to the human driver of a human-driven connected vehicle to cooperatively drive through non-signalized intersections with other CAVs. 

Additionally, a strong assumption has been made in most related studies that, V2X communication performs perfectly in their proposed cooperative driving systems. However, in any wireless communication networks, due to the uncertain reliability of wireless communication links, communication time delays \cite{gao2016robust} and packet losses \cite{ploeg2015graceful} are unavoidable in the information sharing process among connected vehicles. Along with such time-delayed and partially dropped measurements of other connected vehicles' motions, a motion estimation algorithm needs to be designed for the ego vehicle to get a more precise knowledge of others.

Based on the statements above, this study makes the following contributions to this field of research:

\begin{itemize}
    \item A Digital Twin system architecture is built for connected vehicles, where software modules (motion module and communication module) and their algorithms are developed in the digital world, and hardware modules are also specified in the physical world of connected vehicles.
    \item The HMI is designed with augmented reality (AR), where vehicle Digital Twins are visualized to the vehicle driver as reserved slots, assisting the driver to conduct cooperative driving behavior with other connected vehicles at non-signalized intersections.
    \item Unlike many existing studies that only cover one part of the cooperative driving system, we develop an enhanced first-in-first-out (FIFO) slot reservation algorithm to schedule the vehicle sequences, and a consensus motion control algorithm to generate the vehicle speed trajectories. More importantly, we also propose a model-based motion estimation algorithm to tackle the communication issues, including communication delay and packet loss.
    \item Instead of running simplified numerical simulations, we conduct agent-based modeling and simulation of the proposed system with Unity game engine. A real-world corridor with four consecutive non-signalized intersections is built in the simulation environment, where human-in-the-loop (HITL) simulation is conducted with the guidance provided by AR HMI.
\end{itemize}

\subsection{Organization of this Study}
The remainder of this study is organized as follows: Section \ref{sec:Problem} introduces the problem statement and system architecture of this study. Section \ref{sec:Method} proposes the cooperative driving methodology, which includes the enhanced FIFO slot reservation algorithm, the consensus motion control algorithm, and the model-based motion estimation algorithm. Section \ref{sec:AR} introduces the Digital Twin-assisted AR HMI to guide the driver for cooperative driving at non-signalized intersections. Section \ref{sec:ABMS} conducts the agent-based modeling and simulation of the proposed system with Unity game engine, where HITL simulation results are evaluated and sensitivity analysis is also conducted. Finally, the study is concluded with some future directions in section \ref{sec:Con}.

\begin{figure*}[ht!]
    \centering
    \includegraphics[width=2.0\columnwidth]{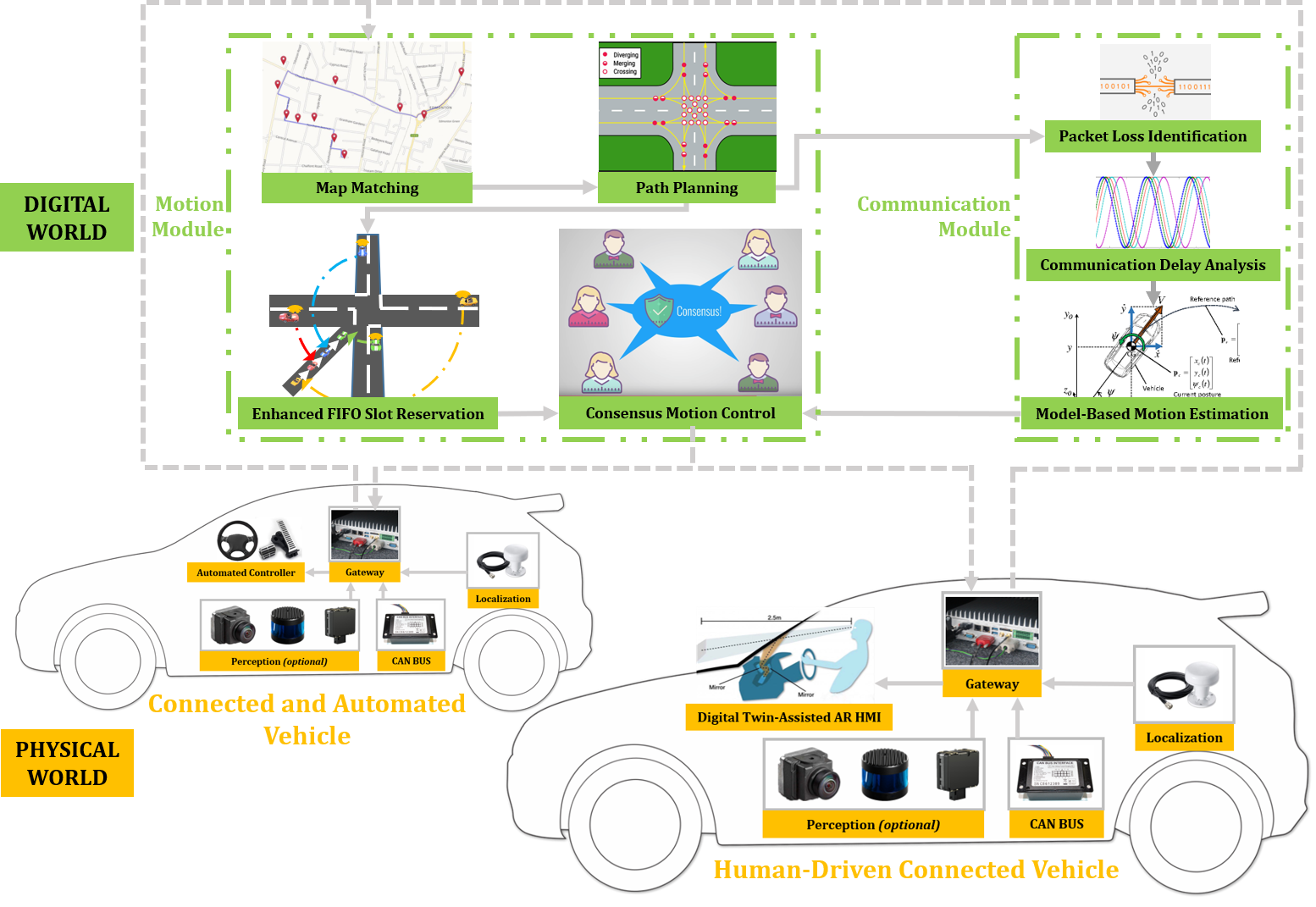}
    \caption{The Digital twin architecture of the proposed cooperative driving system at non-signalized intersections, where the human-driven connected vehicles are equipped with the Digital Twin-assisted HMI}
    \label{Arc}
\end{figure*}

\section{Problem Statement and System Architecture} \label{sec:Problem}
In this study, \first{the Digital Twin technology is adopted to develop} a cooperative driving system that allows connected vehicles to cooperate with each other at non-signalized intersections. Such connected vehicles can either be driven by human drivers with AR HMI, or driven by automated controllers as CAVs. As we adopt the Digital Twin architecture \cite{wang2020adigital}, all connected vehicles in the physical world of this cooperative driving system are connected through the digital world of the system. The proposed cooperative driving system does not specify or require a specific wireless communication technology, which means vehicles can potentially be connected with others at non-signalized intersections through Dedicated Short-Range Communications \cite{kenney2011dedicated}, Cellular Vehicle-to-Everything (C-V2X) \cite{chen2017vehicle}, or a combination of both.

%In this study, the research focus falls into two major parts: 1) cooperative driving methodology at non-signalized intersections (including motion module and communication module), and 2) AR HMI for ego vehicle drivers (designed as Digital Twin representations of other connected vehicles). Specifically, the cooperative driving methodology gathers information from connected vehicles approaching a non-signalized intersection, and computes referenced commands for vehicles to execute. The AR HMI, on the other hand, transfers the referenced commands into visualized assistance for drivers of connected and non-automated vehicles, so they can drive their vehicles to cooperate with other connected vehicles. The general architecture of the proposed system can be illustrated as Fig. \ref{Arc}.

As can be seen from the Digital Twin architecture Fig. \ref{Arc}, there are two major modules in the digital world of the system: the motion module and the communication module. The motion module streamlines the cooperative driving maneuvers with four components: map matching, path planning, enhanced FIFO slot reservation, and consensus motion control. The communication module, on the other hand, analyzes the communication performance among multiple crossing vehicles at an intersection, and then applies the model-based motion estimation algorithm to tackle the communication issues. The estimated motions of target vehicles are also considered as inputs to the motion control component of the motion module. All components in the digital world of the system are discussed in the next section with greater details.

The components in the physical world of this system are all equipped on the connected vehicles, including localization, CAN BUS (for motion measurement), perception (optional as additional information sources to V2X communication), gateway (for data transmission), and \first{the} Digital Twin-assisted AR HMI (for visualized driver assistance). According to the focus of this study, the former four components will not be introduced with more details, where their functionalities in a cooperative driving system can be referred to our previous study \cite{wang2020asurvey}. The Digital Twin-assisted AR HMI is proposed later in Section \ref{sec:AR}. 

\section{Cooperative Driving Methodology Considering Communication Issues} \label{sec:Method}
In this section, the cooperative driving methodology at non-signalized intersections is proposed, which runs in the digital world of the system and allows connected vehicles to cooperate with each other even under the presences of communication delay and packet loss. As can be seen from Fig. \ref{Arc}, the first two components in the motion module are map matching and path planning.

With regards to map matching in this system, its main functionalities are position synchronization and geo-fencing. A pre-built map is stored in the digital world, with information such as intersection geometry, road type, road length and width, link ID and lane ID, waypoints, direction, road speed limit, and etc. At each time step, connected vehicles’ coordinates (i.e., longitude, latitude, and altitude) received from the GNSS are synchronized to the pre-built map, and then the geo-fencing function can be applied to verify whether the other components of the motion module will be applied based on vehicles' positions and conditions.

Regarding to path planning, it generates the lane-level path for vehicles to pass the intersection based on their origins and destinations. As can be seen in the intersection geometry illustrated in Fig. \ref{Arc} above the ``Path Planning'' sign, there are 16 crossing conflicts, 8 merging conflicts, and 8 diverging conflicts at a four-leg single-lane intersection. Therefore, once the paths of different crossing vehicles are generated, the ego vehicle can identify its target vehicle if there exists a conflict point along their paths. Furthermore, based on the intersection geometry, each vehicle's distances to different conflict points (with respect to different target vehicles) can also be calculated accordingly. These information will be fed into the downstream of the motion module, as well as the communication module.

For the sake of simplicity, the technical details of the map matching component and the path planning component are not covered in this study. However, they both are implemented in our agent based modeling and simulation, which is covered in section \ref{sec:ABMS}.

\subsection{Enhanced FIFO Slot Reservation}

The slot reservation component in the digital world of the system takes as inputs the current status of approaching vehicles and their planned paths through the intersection. Then, it schedules the sequences of vehicles to cross the intersection by querying the slot pool. Once the slots are reserved, the ego vehicle can connect and cooperate with its target vehicles that have conflicting paths with itself based on the motion control component and motion estimation component.

\begin{figure}[ht!]%
\centering
\subfigure[Acceleration profile 1]{%
\includegraphics[height=1.1in]{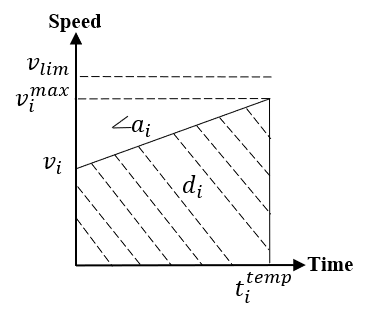}}%
\qquad
\subfigure[Acceleration profile 2]{%
\includegraphics[height=1.1in]{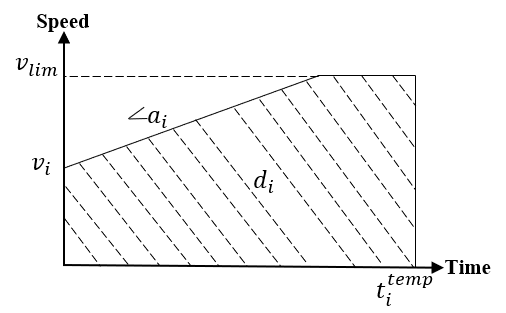}}%
\caption{Two different acceleration profiles: a) vehicle $i$ cannot accelerate to the speed limit $v_{lim}$, and b) vehicle $i$ can accelerate to the speed limit $v_{lim}$}
\label{fig:acc}%
\end{figure}

Different from the traditional first-in-first-out (FIFO) or first-come-first-served (FCFS) policy \cite{dresner2008amultiagent}, which simply assigns a vehicle with a slot when it enters a pre-defined geo-fence, we develop an enhanced FIFO slot reservation algorithm that accounts for the estimated time of arrival (ETA) as well. ETA quantifies a specific period of time $t_i$ that vehicle $i$ is estimated to travel from its current location to a specific conflict point at the intersection. If we assume vehicles will cross the intersection at the intersection speed limit $v_{lim}$, and vehicle $i$'s preferred acceleration $a_i$ is known, we can divide vehicle $i$'s approaching scenario into two categories:

\begin{itemize}
    \item If vehicle $i$ already reaches the speed limit $v_{lim}$ while approaching: Vehicle $i$ will cruise to the conflict point at the intersection, and the temporary ETA $t_{i}^{temp}$ is given by

    \begin{equation}
    t_{i}^{temp}=\frac{d_{i}}{v_{i}}
    \label{eq:t_temp1}
    \end{equation}
    
    \noindent where $d_i$ is the longitudinal distance from vehicle $i$'s current position to the conflict point at the intersection, and $v_i$ is the current longitudinal speed of vehicle $i$.
    
    \item If vehicle $i$'s current longitudinal speed is slower than the speed limit $v_{lim}$ while approaching: Two subcategories will be further defined: a) Vehicle $i$ cannot accelerate to $v_{lim}$ given its preferred acceleration $a_i$ and its current distance to the conflict point at the intersection $d_i$; b) Vehicle $i$ can accelerate to $v_{lim}$. Specifically, the estimated speed profiles of these two subcategories can be visualized as Fig. \ref{fig:acc}, and the temporary ETA $t_{i}^{temp}$ can be given accordingly as below

    \begin{equation}
    \left\{\begin{aligned}
    t_{i}^{temp} &=\frac{-v_{i}+\sqrt{v_{i}^{2}+2 a_{i} d_{i}}}{a_{i}}, \text {if profile 1}\\
    t_{i}^{temp} &=\frac{2 a_{\max } d_{i}+\left(v_{lim}-v_{i}\right)^{2}}{2 a_{\max } v_{lim}}, \text {if profile 2}
    \end{aligned}\right.
    \label{eq:t_temp2}
    \end{equation}
\end{itemize}

%Algorithm 1
\begin{algorithm}
\small
  
\SetAlgoLined
\KwData{Ego vehicle $i$'s path $p_i$ from the current link to the next link, $i$'s distances to conflict points $d_i$, $i$'s longitudinal position $r_i$, $i$'s longitudinal speed $v_i$, $i$'s longitudinal acceleration $a_i$, car-following time headway $t_h$, reservation-trigger time constant $t_\theta$, reservation-trigger geo-fence distance constant $d_\theta$}
\KwResult{Ego vehicle $i$'s reserved slot $s_i$, $i$'s target vehicles $j$}
Ego vehicle $i$ enters the current link\;

\While{$i$ is not assigned $s_i$ at the current intersection}{
Calculate temporary ETA $t_{i}^{temp}$ based on Eq. (\ref{eq:t_temp1}) and (\ref{eq:t_temp2});

\uIf{There is/will be an immediate preceding vehicle $k$ on the same lane of vehicle $i$}{
Update $t_i$ based on Eq. (\ref{eq:ETA});
}
\Else{$t_i=t_{i}^{temp}$;}

Calculate $d_i$ based on $p_i$, $r_i$ and intersection geometry;

\If{$t_i <= t_\theta$ $||$ $d_i <= d_\theta$}{
Query conflicting vehicle $j$ whose path\newline $p_j \cap p_i$ $!=$ $\emptyset$ \;
Query the slot pool regarding the maximum slot number of all conflicting vehicles $s_j^{max}$\;
Assign slot number to the ego vehicle $s_i = s_j^{max} + 1$\;
Connect with conflicting vehicle $j$ as $i$'s target vehicle, and execute motion control and motion estimation algorithms;
}
}
\While{$i$ leaves the current link}{
Reset the slot number $s_i = 0$\;
Disconnect all target vehicles\;}

\caption{\small{Enhanced FIFO slot reservation algorithm for crossing vehicles at a non-signalized intersection}}
\end{algorithm}

Based on either of aforementioned two categories, an additional step in our slot reservation algorithm is that, the temporary ETA value of vehicle $i$ is further updated by the ETA value of its immediate preceding vehicle $k$. This means the traffic condition is also considered in the calculation of ETA, since a following vehicle's ETA cannot be earlier than its preceding vehicle's ETA, where a constant car-following time headway $t_\text{headway}$ must be guaranteed as the delay in between. Therefore, the ETA of vehicle $i$ is finally given as

\begin{equation}
t_{i}=\max\big(t_{i}^{temp},t_{k}+t_\text{headway}\big)
\label{eq:ETA}
\end{equation}

The enhanced FIFO slot reservation algorithm can then be proposed as \textbf{\textit{Algorithm 1}}. As can be seen from line 10 of \textbf{\textit{Algorithm 1}}, the FIFO condition that triggers vehicle $i$'s slot reservation request is both time-based and location-based: Either its ETA is lower than a predefined time constant $t_\theta$, or it enters a pre-defined geo-fence of this intersection. This FIFO enhancement prevents some corner cases such as vehicle $i$ enters the geo-fence earlier, but has a much lower speed (i.e., expected to arrive at the intersection later) than its conflicting vehicle $j$ coming from another leg of the intersection. In that case, the conflicting vehicle $j$ needs to significantly decelerate to follow vehicle $i$ to cross the intersection, which dampens overall traffic throughput and energy efficiency.

\subsection{Consensus Motion Control}
%Once a connected vehicle is assigned a slot and connected with its target vehicles, vehicle information is constantly transmitted among them. The control module of the ego vehicle aims to adjust the positions and sizes of its target vehicles' reserved slots, so the slots can be better visualized for the AR HMI.

%First, a target speed of the ego vehicle is calculated, based on the information received from its leading target vehicle $j$, whose reserved slot is right in front of the ego vehicle's slot (i.e., $s_j = s_i - 1$). This target speed allows the ego vehicle to follow the movement of its target vehicle's slot with the car-following time headway. A feedforward/feedback control algorithm is developed to calculate this target speed (that needs to be executed at the next time step) $v_i(t + \delta t)$. The feedback consensus control part is written as follow

The motion control component in the digital world of the system takes as input the assigned slots and the target vehicles of the ego vehicle, and then computes its referenced motion. First, given the longitudinal dynamics of a vehicle $i$ as the following equations:

\begin{equation} \label{eq:motion}
\begin{split}
\dot{d}_{i}(t) &=v_{i}(t) \\
\dot{v}_{i}(t) &=a_{i}(t) \\
a_{i}(t) &=\frac{1}{m_i}[F_{n e t_{i}}(t)-G_{i} T_{b r_{i}}(t)-c_{v i} v_{i}(t)^{2}\\
&-c_{f i} v_{i}(t)-f_{drag_{i}}(t)]
\end{split}
\end{equation}

\noindent where ${d}_{i}(t)$, $v_{i}(t)$, and $a_{i}(t)$ denote the longitudinal distance to the conflict point, longitudinal speed, and longitudinal acceleration of vehicle $i$ at time $t$, respectively; $m_i$ denotes the mass of vehicle $i$; $F_{n e t_{i}}(t)$ denotes the net engine force of vehicle $i$ at time $t$, which mainly depends on the vehicle speed and the throttle angle; $G_i$ denotes the effective gear ratio from the engine to the wheel of vehicle $i$; $T_{b r_{i}}(t)$ denotes the brake torque of vehicle $i$ at time $t$; $c_{vi}$ denotes the coefficient of aerodynamic drag of vehicle $i$; $c_{fi}$ denotes the coefficient of friction force of vehicle $i$; $f_{drag_{i}} (t)$ denotes the mechanical drag of vehicle $i$ at time $t$.

We can then derive the following equations from the principle of vehicle dynamics when the braking maneuver is deactivated, i.e., vehicle $i$ is accelerating by the net engine force:

\begin{equation}
F_{n e t_{i}}(t)=\ddot{d}_{i}(t) m_{i}+c_{v i} \dot{d}_{i}(t)^{2}+c_{p i} \dot{d}_{i}(t)+f_{drag_{i}}(t)
\end{equation}

\noindent and we have the following equation when the braking maneuver is activated, i.e., vehicle $i$ decelerates by the brake torque:

\begin{equation}
T_{b r_{i}}(t)=\frac{\dot{d}_{i}(t) m_{i}+c_{v i} \dot{d}_{i}(t)^{2}+c_{p i} \dot{d}_{i}(t)+f_{drag_{i}}(t)}{G_{i}}
\end{equation}

Note that the net engine force is a function of the vehicle speed and the throttle angle, which is typically based on the steady-state characteristics of engine and transmission systems. The associated mathematical derivations can be referred to \cite{xiao2011practical}.

Based on the existing literature \cite{milanes2014cooperative}, the motion control of a connected vehicle is based on a hierarchical strategy, where the high-level controller generates a referenced motion (the first two equations in equation (\ref{eq:motion})), while the low-level controller commands the vehicle actuators to track the referenced motion (the last equation in equation (\ref{eq:motion})). In this study, the motion control component in the digital world only focuses on the high-level control part, aiming to generate the referenced motion of the ego vehicle based on the inputs from its target vehicles.

\begin{figure}[!htb]
\centering
\includegraphics[width=0.5\textwidth]{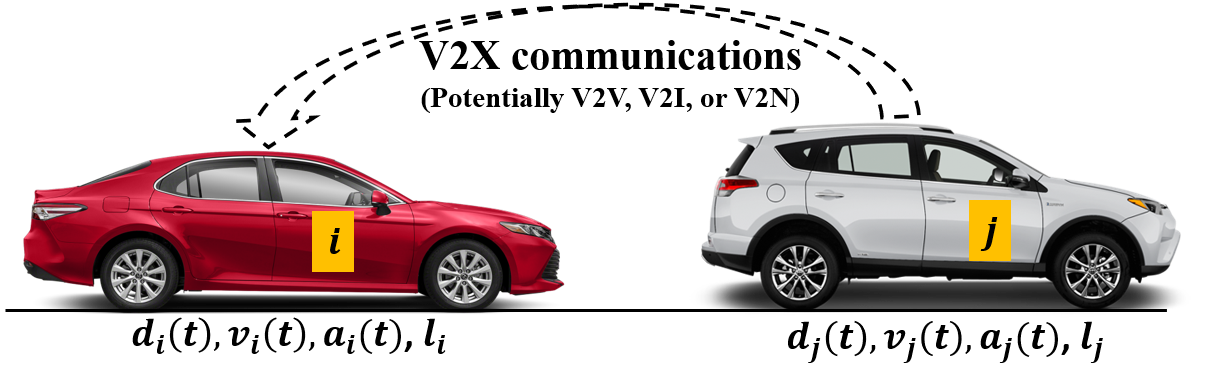}
\caption{The illustration of vehicle parameters in a V2X communication environment with an ego vehicle $i$ and its target vehicle $j$}
\label{fig:dynamics}
\end{figure}

Given a car-following scenario shown as Fig. \ref{fig:dynamics}, where the ego vehicle $i$ gets information from its target vehicle $j$ through V2X (e.g., V2V, V2I, or V2N) communication, the motion consensus can be generalized as a longitudinal control problem as follows:

\begin{equation} \label{eq:converge}
\begin{array}{l}
d_{i}(t) \rightarrow d_{j}(t) - d_{\text{headway}} \\
v_{i}(t) \rightarrow v_{j}(t) \\
a_{i}(t) \rightarrow a_{j}(t)
\end{array}
\end{equation}

\noindent where $d_{\text{headway}}$ is the desired distance headway between two vehicles. In order to achieve the motion consensus in equation (\ref{eq:converge}), a double-integrator consensus-based longitudinal motion control algorithm \cite{wang2019lookup} can be given as:

\begin{equation} \label{eq:consensus}
\begin{split}
\dot{d}_{i}(t) &=v_{i}(t) \\
\dot{v}_{i}(t) &=-\alpha_{i j} k_{i j} \cdot[(d_{i}(t)-d_{j}(t)+l_{j}+v_{i}(t) \cdot t_{i j}^{g}(t))\\
&+\gamma_{i}\cdot(v_{i}(t)-v_{j}(t))]
\end{split}
\end{equation}

\noindent where $\alpha_{i j}$ is the adjacency matrix of the directed graph (i.e., V2X communication topology between vehicle $i$ and $j$); The term $[l_{j}+v_{i}(t) \cdot t_{i j}^{g}(t)]$ is another form of the term $d_{\text{headway}}$ in equation (\ref{eq:converge}), where $t_{i j}^{g}(t)$ is the time-variant desired time gap between two vehicles, which can be adjusted by many factors like road grade, vehicle mass, braking ability, and etc.; The control gains $k_{ij}$ and $\gamma_i$ in this algorithm are tuned by a feedforward control algorithm to guarantee the safety, efficiency, and comfort of this slot-following process, which can be summarized as:

\begin{equation}
    \{k_{ij}, \gamma_i\} = f\big(v_i(0), v_j(0), d_i(0) - d_j(0)\big)
\end{equation}

\noindent A lookup-table can be constructed to dynamically calculate these two control gains based on the initial speeds of two vehicles, as well as their initial headway. The details of this feedforward control algorithm can be referred to our previous work \cite{wang2019lookup}. With this longitudinal motion control algorithm equation (\ref{eq:consensus}), vehicle $i$ in Fig. \ref{fig:dynamics} is able to converge its longitudinal speed $v_{i}(t)$ to vehicle $j$'s longitudinal speed $v_{j}(t)$, and converge its longitudinal distance ${d}_{i}(t)$ to vehicle $j$'s longitudinal distance ${d}_{j}(t)$ (with a difference in between).

However, the aforementioned algorithm does not consider the communication delay issue. It is without a doubt that, whatever information vehicle $i$ receives is not exactly the current information of vehicle $j$, due to the unavoidable transmission time $\tau_{ij}(t)$. Therefore, equation (\ref{eq:consensus}) can be further written into the following form while considering the time-variant communication delay:

\begin{equation} \label{eq:consensusdelay}
\begin{split}
\dot{d}_{i}(t) &=v_{i}(t) \\
\dot{v}_{i}(t) &=-\alpha_{i j} k_{i j} \cdot[(d_{i}(t)-d_{j}(t-\tau_{ij}(t))+l_{j}+v_{i}(t) \cdot t_{i j}^{g}(t)))\\
&+\gamma_{i}\cdot(v_{i}(t)-v_{j}(t-\tau_{ij}(t)))]
\end{split}
\end{equation}

\noindent where the referenced acceleration $\dot{v}_{i}(t)$ of vehicle $i$ can be considered as the referenced motion of the vehicle. However, in order to tackle the communication issues of time delay and packet loss, a model-based motion estimation algorithm is proposed to adjust the referenced motion, and then pass to the physical world of the system for execution.

\subsection{Model-Based Motion Estimation for Communication Delay and Packet Loss}

Time delay and packet loss are two major issues that impair the performance of V2X communication in connected vehicle applications. Different from most of the existing literature that tackle these two issues separately \cite{gao2016robust, ploeg2015graceful}, we propose a model-based motion estimation algorithm in this study to address them at the same time, allowing connected vehicles to conduct cooperative driving maneuver at non-signalized intersections with moderate communication delay and packet loss.

As stated below, \textit{\textbf{Algorithm 2}} is the main function of the proposed motion estimation algorithm, where \textit{\textbf{Algorithm 2.1}}, \textit{\textbf{Algorithm 2.2}}, and \textit{\textbf{Algorithm 2.3}} are called in this main function.

\begin{algorithm}
\small
  
\SetAlgoLined
\KwResult{Vehicle $j$'s estimated longitudinal motion $\mathbf{\Tilde{V}_j(t)}$ and $\mathbf{\Tilde{D}_j(t)}$ in the future horizon $[t+1, t+N]$}

Vehicle $i$ associates with its target vehicle $j$\;

\While{communication between vehicle $i$ and $j$ is currently on}{

\uIf{$n=j==0$, namely vehicle $j$ is the leader of a communication topology and does not have any target vehicle}
{Vehicle $j$ estimate its future longitudinal speed trajectory $\mathbf{\Tilde{V}_j(t)}$ based on \textbf{\textit{Algorithm 2.1}}\;
Vehicle $j$ cumulatively estimates its future longitudinal distance trajectory $\mathbf{\Tilde{D}_j(t)}$ based on \textbf{\textit{Algorithm 2.2}};
}

\Else{
\For{$n==1 \rightarrow j$}{

% Communication Delay
%\uIf{Communication delay $\tau_{n(n-1)}(t+k)$ is less than the duration of a prediction time step $\delta t$}
%{$\Tilde{v}_{n-1}(t+k)=\Tilde{v}_{n-1}(t+k-\tau_{n(n-1)}(t+k)$\;
%$\Tilde{d}_{n-1}(t+k)=\Tilde{d}_{n-1}(t+k-1) + \Tilde{v}_{n-1}(t+k) \cdot \tau_{n(n-1)}(t+k)$;}
%\Else{Consider delay as packet loss, where $\Bar{k}=\tau_{n(n-1)}(t+k)/\delta t$;}

% Packet Loss
\uIf{Vehicle $n$ is connected to its target vehicle $n-1$, namely no packet loss is in presence at this time step $t$}
{Vehicle $n$ estimates its future longitudinal speed trajectory $\mathbf{\Tilde{V}_n(t)}$ based on $\mathbf{\Tilde{V}_{n-1}(t)}$ and $\mathbf{\Tilde{D}_{n-1}(t)}$ with \textbf{\textit{Algorithm 2.3}}\;}
\Else{Vehicle $n$'s future longitudinal speed trajectory estimate stays the same since no information update  $\mathbf{\Tilde{V}_n(t)} = \mathbf{\Tilde{V}_n(t-1)}$\;}

Vehicle $n$ cumulatively estimates its future longitudinal distance trajectory $\mathbf{\Tilde{D}_n(t)}$ based on $\mathbf{\Tilde{V}_n(t)}$ with \textbf{\textit{Algorithm 2.2}}\;}
Vehicle $j$'s estimated motion $\mathbf{\Tilde{V}_j(t)}$ and $\mathbf{\Tilde{D}_j(t)}$ can be derived when $n==j$;
}

Vehicle $j$ sends $\mathbf{\Tilde{V}_j(t)}$ and $\mathbf{\Tilde{D}_j(t)}$ to its following vehicle $i$;
}

\While{communication between vehicle $i$ and $j$ is currently off due to packet loss}
{
\uIf{Packet loss period exceeds the safety threshold}
{Trigger the fail-safe plan and transfer the upcoming non-signalized intersection into an all-way stop;}
\Else{At time step $(t+k) \subseteq [t+1,t+N]$, vehicle $i$ extras $\Tilde{v}_j(t+k)$ from $\mathbf{\Tilde{V}_j(t)}$, and $\Tilde{d}_j(t+k)$ from $\mathbf{\Tilde{D}_j(t)}$, and uses them as the inputs of the motion control algorithm;}
}

Vehicle $i$ disassociates with its target vehicle $j$;

\caption{\small{Main function of the motion estimation algorithm, with three algorithms being called inside}}
\end{algorithm}

%Algorithm 2.1
\textit{\textbf{Algorithm 2.1:}} This algorithm estimates vehicle $j$'s future longitudinal speed trajectory when it does not have any target vehicle to follow. In this case, vehicle $j$ will converge to its target speed $v_j(t \rightarrow \infty)$, which is a known and preset value. Specifically, this algorithm is proposed based upon the Intelligent Driver Model (IDM) \cite{treiber2000congested}, where the estimated speed of vehicle $j$ at time step $(t+k)$ is given as follows:

\begin{equation}
\Tilde{v}_j(t+k)=\Tilde{v}_j(t+k-1) + a_j^{\max } \cdot\left[1-\left(\frac{\Tilde{v}_j(t+k-1)}{v_j(t \rightarrow \infty)}\right)^{\sigma}\right] \cdot \delta t
\end{equation}

\noindent where $k \subseteq [1, N]$, and $\Tilde{v}_j(t) = v_j(t)$; $a_j^{max}$ is a preset constant denoting vehicle $j$'s maximum changing rate of longitudinal speed; $\sigma$ is the free acceleration exponent defined by IDM, which characterizes how the acceleration of the vehicle decreases with speed, namely, $\sigma = 1$ corresponds to a linear decrease, while $\sigma \rightarrow \infty$ leads to a constant acceleration; $\delta t$ is the duration of a prediction time step. Based on this algorithm, the future longitudinal speed trajectory of vehicle $j$ can be given as $\mathbf{\Tilde{V}_j(t)}=\Big(\Tilde{v}_j(t+1), \Tilde{v}_j(t+2), ..., \Tilde{v}_j(t+k), ..., \Tilde{v}_j(t+N)\Big)$.

%Algorithm 2.2
\textit{\textbf{Algorithm 2.2:}} This algorithm estimates vehicle $n$'s future longitudinal distance trajectory $\mathbf{\Tilde{D}_n(t)}$ based on its estimated longitudinal speed trajectory $\mathbf{\Tilde{V}_n(t)}$, which is given as:

\begin{equation}
\Tilde{d}_n(t+k)= \Tilde{d}_n(t+k-1) + \Tilde{v}_n(t+k-1) \cdot \delta t
\end{equation}

\noindent where $k \subseteq [1, N]$, and $\Tilde{d}_n(t) = d_n(t)$. Based on this algorithm, the future longitudinal position trajectory of vehicle $n$ can be given as $\mathbf{\Tilde{D}_n(t)}=\Big(\Tilde{d}_n(t+1), \Tilde{d}_n(t+2), ..., \Tilde{d}_n(t+k), ..., \Tilde{d}_n(t+N)\Big)$.

%Algorithm 2.3
\textit{\textbf{Algorithm 2.3:}} This algorithm estimates vehicle $n$'s future longitudinal speed trajectory when it has a target vehicle $(n-1)$ to follow, meanwhile also considers the presence of communication delay $\tau_{n(n-1)}(t+k)$:

\begin{itemize}
    \item When $\tau_{n(n-1)}(t+k)<\delta t$, namely the communication delay is less than the duration of a prediction time step, the target vehicle $(n-1)$'s longitudinal speed is assumed unchanged during this delayed period:

        \begin{equation}
        \Tilde{v}_{n-1}(t+k)=\Tilde{v}_{n-1}\Big(t+k-\tau_{n(n-1)}(t+k)\Big)
        \end{equation}
    
    \item When $\tau_{n(n-1)}(t+k)>=\delta t$, namely the communication delay is equal to or longer than the duration of a prediction time step, then

        \begin{equation}
        \begin{split}
        \Tilde{v}_{n-1}(t+k) = \Tilde{v}_{n-1}\Big(t+k-\tau_{n(n-1)}(t+k)\Big) \\
         +\dfrac{\tau_{n(n-1)}(t+k)}{\delta t}\cdot\dot{\Tilde{v}}_{n-1}\Big(t+k-\tau_{n(n-1)}(t+k)\Big)
        \end{split}
        \end{equation}
\end{itemize}

In either case, target vehicle $(n-1)$'s longitudinal distance can be adjusted by:

\begin{equation}
\Tilde{d}_{n-1}(t+k)=\Tilde{d}_{n-1}(t+k-1) + \Tilde{v}_{n-1}(t+k) \cdot \tau_{n(n-1)}(t+k)
\end{equation}

Then, vehicle $(n-1)$'s future longitudinal motion at each time step is used to estimate vehicle $n$'s future longitudinal speed as:

\begin{equation} \label{algorithm4}
\begin{split}
\Tilde{v}_{n}(t+k) & =\Tilde{v}_{n}(t+k-1)-\alpha_{n (n-1)} k_{n (n-1)} \\ 
& \cdot\Bigg[\bigg(\Tilde{d}_{n}(t+k)-\Tilde{d}_{n-1}\Big(t+k\Big) \\
& +l_{n-1}+\Tilde{v}_{n}(t+k) \cdot t_{n (n-1)}^{g}(t+k)\Big)\bigg) \\
& +\gamma_{n}\cdot\bigg(\Tilde{v}_{n}(t+k)-\Tilde{v}_{n-1}(t+k\Big)\bigg)\Bigg]
\end{split}
\end{equation}

Parameters of this algorithm are set according to the consensus motion control algorithm equation (\ref{eq:consensus}). Based on this algorithm, the future longitudinal speed trajectory of vehicle $n$ can be given as $\mathbf{\Tilde{V}_n(t)}=\Big(\Tilde{v}_n(t+1), \Tilde{v}_n(t+2), ..., \Tilde{v}_n(t+k), ..., \Tilde{v}_n(t+N)\Big)$.

\section{Digital Twin-Assisted AR HMI} \label{sec:AR}
In this section, we propose an AR HMI to visualize the Digital Twins of target vehicles based on the slot reservation, motion control, and motion estimation algorithms proposed in the previous section. This Digital Twin-assisted AR HMI can provide the guidance to the driver of a human-driven connected vehicle to cooperatively drive through non-signalized intersections with other connected vehicles (both human-driven ones and CAVs). \first{It should be noted that, this AR HMI is an optional module of the proposed Digital Twin-assisted cooperative driving system, which shows one possible HMI design among others for human-driven connected vehicles. The challenges brought by this AR HMI design, such as user compliance, fall outside of the scope of this study.}

\begin{figure}[ht!]
    \centering
    \includegraphics[width=1.0\columnwidth]{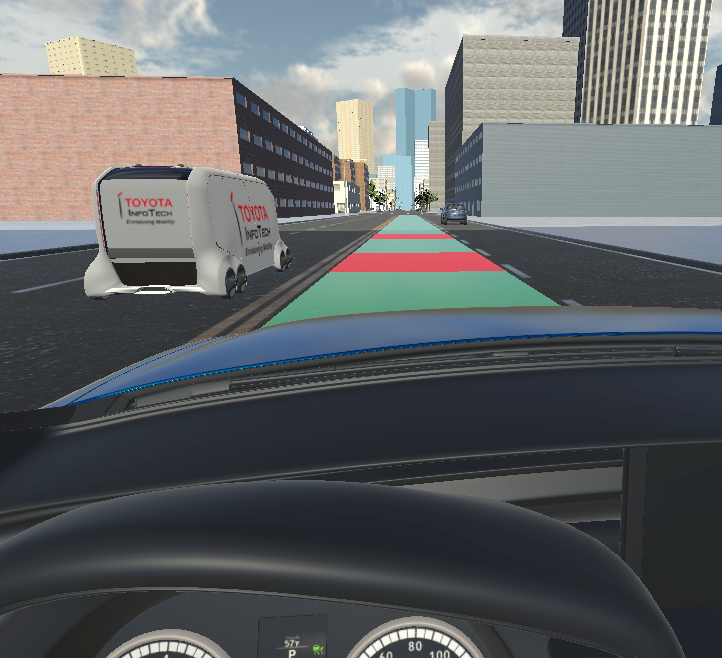}
    \caption{Driver's field-of-view of the AR HMI, where those two red slots stand for the Digital Twins of two target vehicles approaching from other legs of this intersection}
    \label{AR2}
\end{figure}

The design of the Digital Twin-assisted AR HMI from the driver's field-of-view is illustrated in Fig. \ref{AR2}. The ego vehicle is approaching a non-signalized intersection, where the two red slots are \first{the} Digital Twins of two target vehicles approaching from other legs of this intersection, reserved by the aforementioned enhanced FIFO slot reservation algorithm. The driver of this ego vehicle is guided to drive the vehicle to keep in the green available slots, so it can avoid collision while crossing the intersection.

Specifically, the slot adjustment algorithm is proposed as \textbf{\textit{Algorithm 3}}, where the positions and sizes in the world referenced frame of the red reserved slots are calculated based on the outputs from the slot reservation, motion control, and motion estimation algorithms. Specifically, the length $l_{s_j}$ of the reserved slot is adjusted by the time gap $t_{i j}^{g}$, which considers the following clearance as a redundancy design of the system. The value of ``2'' in front of the $v_i * t_{i j}^{g}$ term can be tuned based on V2X communication performances or motion estimation confidence of the system, where a higher value means more redundancy.

\begin{algorithm}
\small
\SetAlgoLined
\KwData{Ego vehicle $i$'s longitudinal speed $v_i$, $i$'s lateral position $x_i$, $i$'s target vehicle $j$, $i$'s longitudinal distance to the conflict point $d_i$, $j$'s estimated longitudinal distance to the conflict point $\Tilde{d}_j$, car-following time gap $t_{i j}^{g}$, $j$'s length $l_j$, $j$'s width $w_j$}
\KwResult{$j$'s reserved slot's longitudinal distance to the conflict point $d_{s_j}$, lateral position $x_{s_j}$, length $l_{s_j}$, width $w_{s_j}$}

\For{Ego vehicle $i$'s target vehicle $j$}{

\While{$i$ does not cross the conflict point $O_{ij}$}{
Calculate $j$'s reserved slot's longitudinal distance to the conflict point $d_{s_j} = \Tilde{d}_j$\;
Set $j$'s reserved slot's lateral position $x_{s_j} = x_i$\;
Calculate $j$'s reserved slot's width $w_{s_j} = w_j$\;
Calculate $j$'s reserved slot's length $l_{s_j} = l_j + 2 * v_i * t_{i j}^{g}$\;
}

\While{$i$ crosses the conflict point $O_{ij}$}{
Reset the slot information $d_{s_j}, x_{s_j}, w_{s_j}, l_{s_j}$\;
}
}

\caption{\small{Slot adjustment in the world referenced frame}}
\end{algorithm}

\begin{figure}[ht!]
    \centering
    \includegraphics[width=1.0\columnwidth]{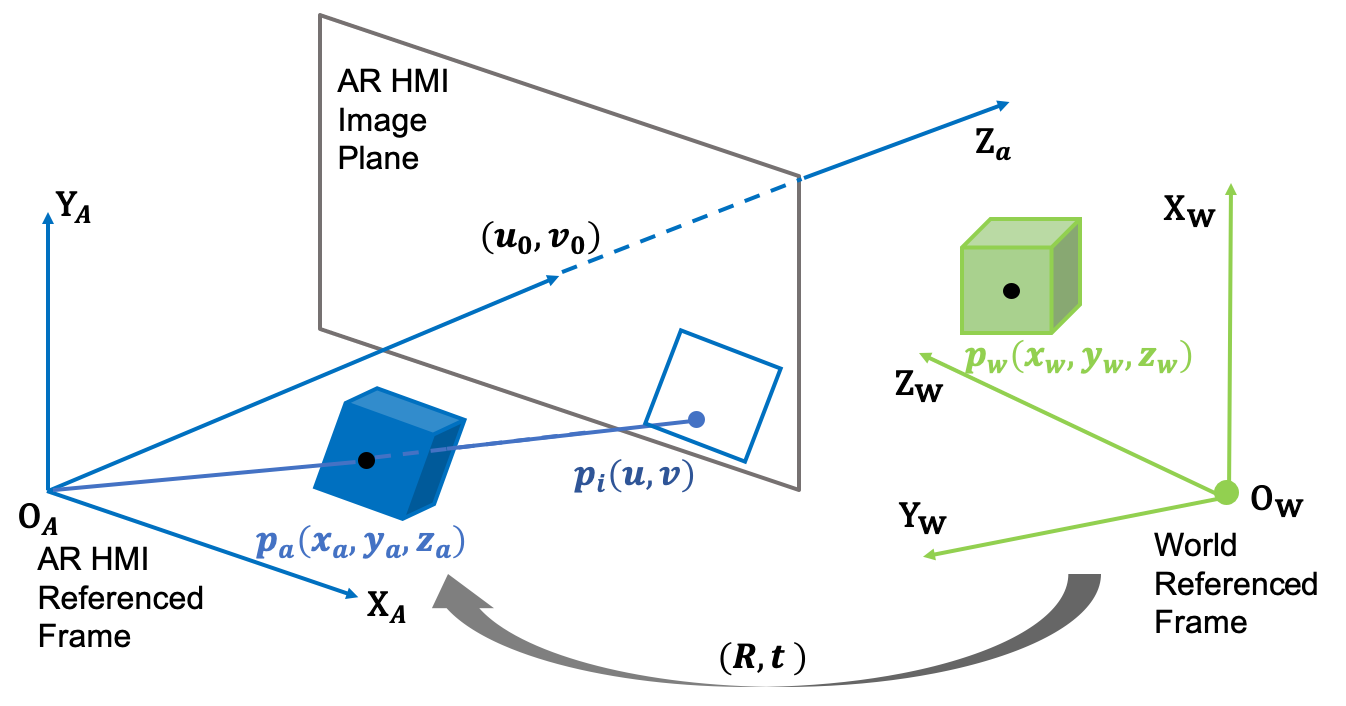}
    \caption{Coordinate transformation of the slot from the world referenced frame to AR HMI referenced frame}
    \label{transform}
\end{figure}

Once the positions and sizes of the slots are calculated in the world referenced frame, the next step is to transform those coordinates into a format that can be visualized on the AR HMI of the vehicle. The AR HMI is expected to be displayed on an image plane, such as on the windshield (i.e., in the format of head-up display), or on an external screen (e.g., vehicle infotainment system or mobile device). A front-view camera is needed to identify the road geometry, so the slots can be correctly overlaid on the road surface from the driver's field-of-view. Therefore, we develop a coordinate transformation algorithm based on the pinhole camera projection model, which is illustrated in Fig. \ref{transform}. 

The extrinsic parameter matrix in this algorithm identifies the transformation between the world referenced frame and the AR HMI referenced frame. It consists of a $3\times3$ rotation matrix $R$ and a $3\times1$ translation vector $t$. Given a 3D point of the slot in the world referenced frame $p_w(x_w, y_w, z_w)$, its corresponding point $p_a$ in the AR referenced frame can be calculated as

\begin{equation}
   p_a = \begin{bmatrix} R & t \end{bmatrix}_{3 \times 4} \begin{pmatrix} x_w\\y_w\\z_w\\1 \end{pmatrix}_{4 \times 1}
\end{equation}

Then, the intrinsic parameter matrix is applied, which contains the parameters of the AR HMI's projection device, such as the focal length and lens distortion. Let ($u_0, v_0$) be the coordinates of the principle point of the image plane (i.e., image center), $d_x$ and $d_y$ be the physical size of pixels, and $f$ be the focal length, the projected point $p_i(u,v)$ on the AR HMI image plane can be calculated as

\begin{equation}
   p_i = 
   \begin{bmatrix} z_a d_x / f & 0 &  -z_a d_x u_0/ f \\
   0 & z_a d_y/ f & -z_a d_y v_0/ f\\
   0 & 0 & z_a \end{bmatrix}_{3 \times 3} \begin{pmatrix} p_a \end{pmatrix}_{3 \times 1}
\end{equation}

Therefore, given the position and size of any slot in the world referenced frame, we are able to transform the slot to the AR HMI image plane, so it can be properly shown to the driver of the vehicle.

\section{Agent-Based Modeling and Simulation using Unity Game Engine} \label{sec:ABMS}
\subsection{Modeling and Simulation Environment in Unity Game Engine}
Game engines enable the design of video games for software developers, which typically consist of a rendering engine for 2-D or 3-D graphics, a physics engine for collision detection and response, and a scene graph for the management of multiple elements (e.g., models, sound, scripting, threading, etc.). Along with the rapid development of game engines in recent years, their functions have been broadened to a wider scope: data visualization, training, medical, and military use. Game engines also become popular options in the development of advanced vehicular technology \cite{ma2020new}, which have been used to study driver behaviors \cite{wang2020driver}, prototype connected vehicle systems \cite{wang2019cooperative, liu2020sensor}, and simulate autonomous driving \cite{dosovitskiy2017carla, rong2020lgsvl}.

\begin{figure}
    \centering
    \includegraphics[width=0.8\columnwidth]{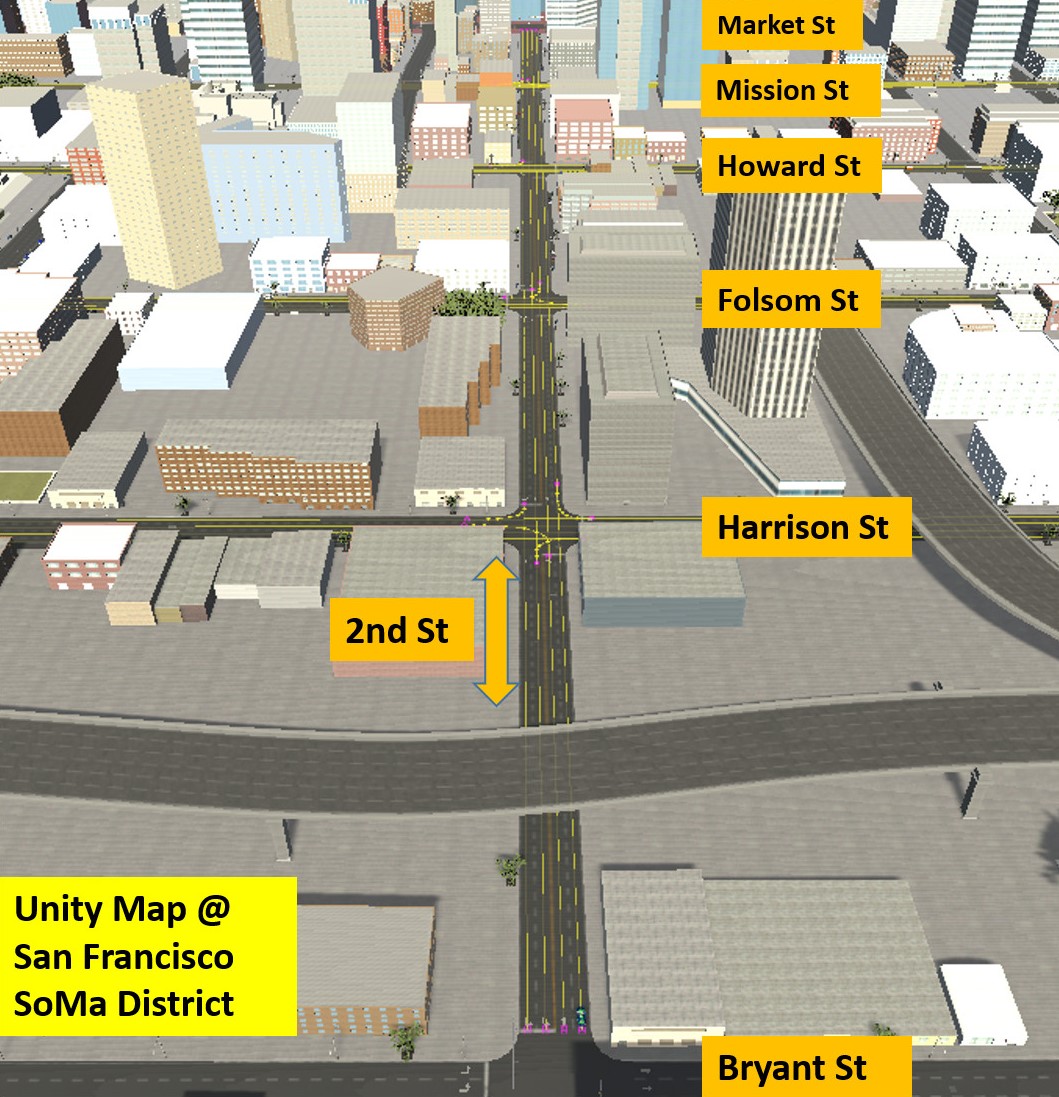}
    \caption{The map with a four-intersection (each with four legs) corridor built in Unity game engine based on the SoMa district in San Francisco}
    \label{map}
\end{figure}

In this study, we adopt Unity game engine to conduct modeling and evaluation of the proposed cooperative driving system, given its advantages of graphics design and visualization, as well as its easiness to connect with external driving simulators \cite{unity}. As shown in Fig. \ref{map}, the map built by LGSVL is adopted in our study, which is based on the South of Market (SoMa) district in San Francisco \cite{rong2020lgsvl}. Shown as yellow lines on the road surface, centimeter-level routes along the 2nd Street, Harrison Street, Folsom Street, Howard Street, and Mission Street are further modeled in this study, so map matching and path planning features can be enabled in our simulation. The slot reservation, motion control, and motion estimation algorithms are modeled for each agent (i.e., connected vehicle) in this environment through Unity's C\# API. The AR HMI is also modeled on the ego vehicle through Unity's visualization feature, which is briefly shown as Fig. \ref{AR2}.

\begin{figure}
    \centering
    \includegraphics[width=0.8\columnwidth]{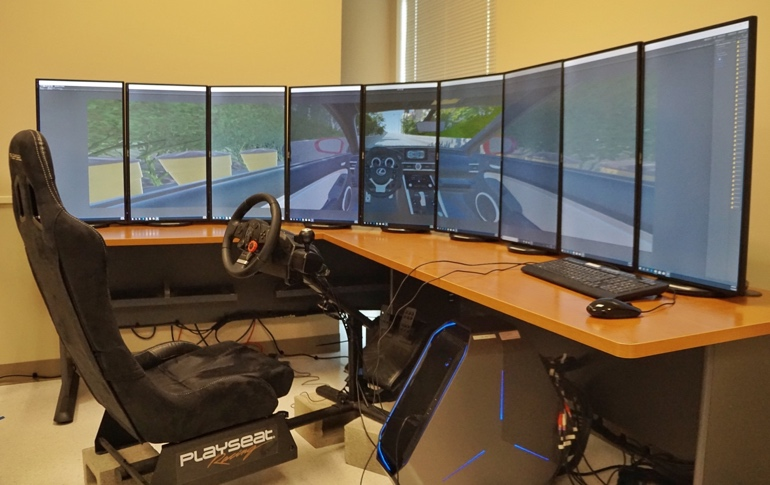}
    \caption{Driving simulator platform to conduct human-in-the-loop (HITL) simulation}
    \label{Sim}
\end{figure}

In this simulation, we set the time-variant communication delay based on the results of our previous field test regarding LTE-based V2X communication \cite{wang2020adigital}. Note that this communication technology normally has higher communication delays than DSRC, so we adopt it in the simulation as a stress test of our motion estimation algorithm. Specifically, we set the communication delay as a normal distribution with a mean value of 40 ms and a standard deviation of 0.0259. 

Additionally, we model the packet loss in the simulation with a hybrid model, which includes both random packet loss over the whole simulation period, and certain packet loss during some specific time periods. The random packet loss is to simulate the relatively periodic communication issues, while the certain  packet loss is to simulate the non-line-of-sight (NLOS) scenario when the communication is obstructed by physical objects in the simulation environment (e.g., tall buildings or bridges). The packet loss is modelled and simulated in a more frequent manner than it is supposed to be in the real world, so we can conduct stress test of our motion estimation algorithm, similar to the communication delay setup.

To evaluate the proposed AR HMI, we conduct HITL simulation with drivers controlling the human-driven connected vehicles through the external driving simulator. As shown in Fig. \ref{Sim}, the driving simulator platform is built with a Windows desktop (processor Intel Core i7-9750 @2.60 GHz, 32.0 GB memory, NVIDIA Quadro RTX 5000 Max-Q graphics card), a Logitech G29 Driving Force racing wheel, and Unity 2019.2.11f1.

The invited participants in this simulation are advised to drive the ego vehicle in the Unity environment. The ego vehicle starts from the Bryant Street with zero speed, and then crosses four consecutive non-signalized intersections along the 2nd Street. All other vehicles in the simulation are non-player characters (NPCs), which run the proposed motion control and estimation algorithms as CAVs. As a comparison, all participants also drive the ego vehicle in the baseline scenario, where all four intersections have traditional fixed-timing traffic signals. NPC vehicles are randomly generated from all legs at each intersection, and they are not enabled with any CAV feature in the baseline scenario.

\subsection{Human-in-the-Loop Simulation Results}
\begin{figure}
    \centering
    \includegraphics[width=1.0\columnwidth]{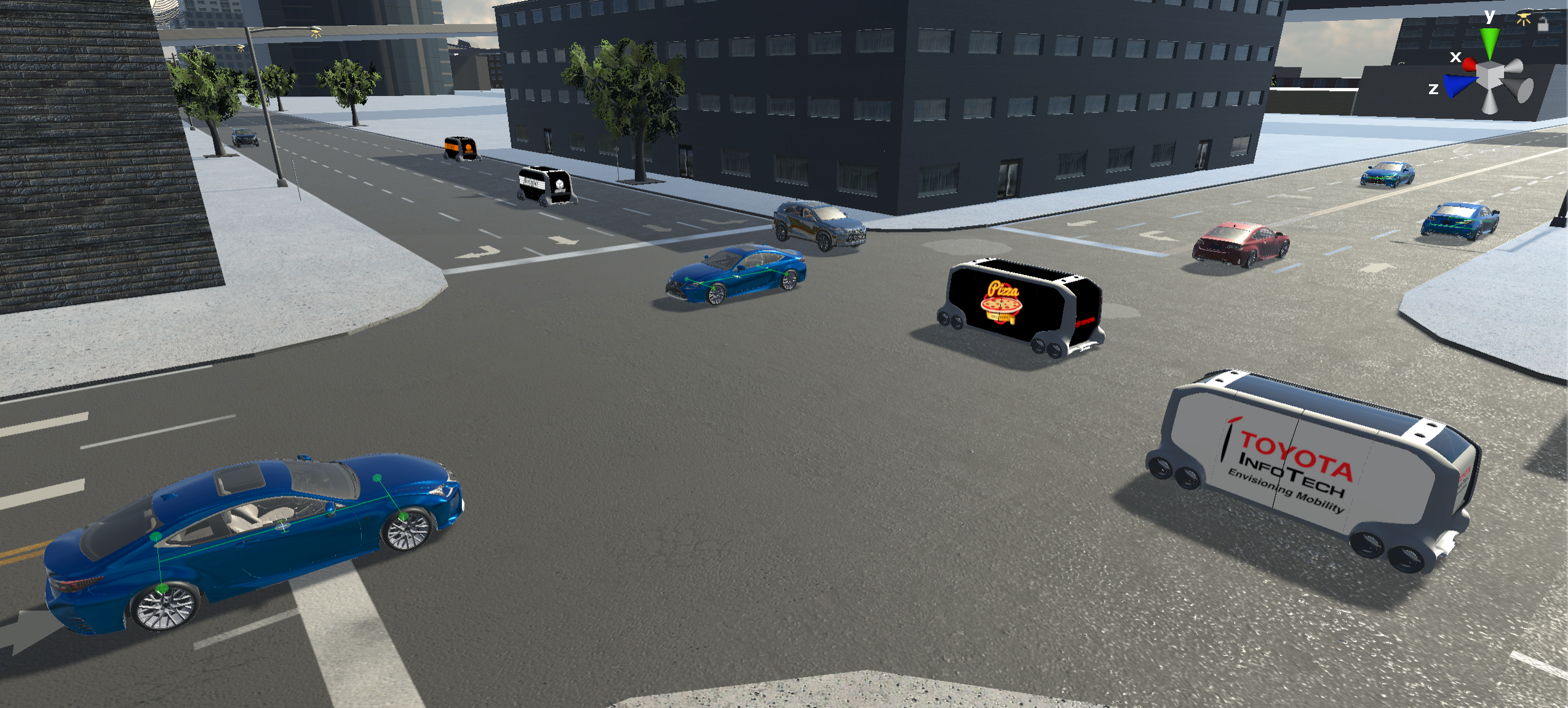}
    \caption{A snapshot of the Unity simulation, where connected vehicles cooperate with each other at a non-signalized intersection}
    \label{unity}
\end{figure}

A snapshot of the Unity simulation is shown as Fig. \ref{unity}, where connected vehicles automatically cooperate with each other to cross this non-signalized intersection without any collision or full stop. As an example, we pick out one of the trips where the participant drives the ego vehicle (human-driven connected vehicle with AR HMI) to cross the first intersection on that corridor. 

\begin{figure}
    \centering
    \subfigure[]{%
    \includegraphics[width=0.8\columnwidth]{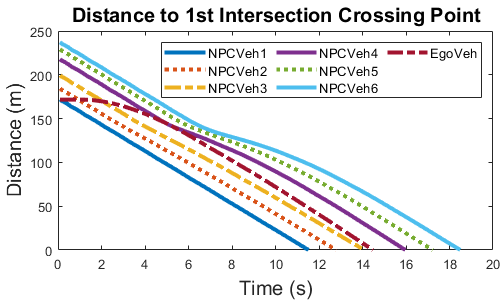}
    \label{fig:distance}}
    \\
    \subfigure[]{%
    \includegraphics[width=0.8\columnwidth]{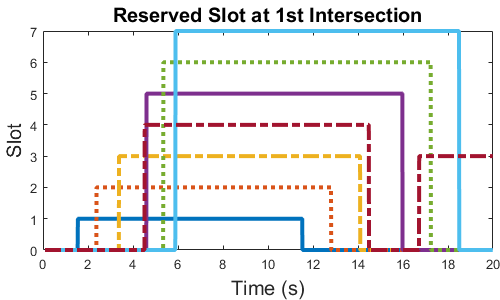}
    \label{fig:slot}}
    \\
    \caption{A sample result of connected vehicles crossing the first intersection in the simulation, where the ego vehicle is driven by a human driver on the driving simulator platform}
    \label{Result}
\end{figure}

The distance-time plot in Fig. \ref{fig:distance} shows the ego vehicle is able to keep a relatively safe distance regarding its target vehicles, including its immediate preceding NPC vehicle 3. Meanwhile, the ego vehicle also acts as a target vehicle for NPC vehicle 4, 5 and 6, where NPC vehicle 4 considers the ego vehicle as its immediate preceding vehicle. Since the proposed motion control and estimation algorithms are applied to these three NPC vehicles, they consecutively decelerated during 5-12 s to maintain a relatively safe distance with their immediate preceding vehicles.

The process of all seven vehicles reserving slots is shown in Fig. \ref{fig:slot}, which corresponds to the vehicle trajectories in Fig. \ref{fig:distance}. Once a vehicle is assigned a slot by our proposed enhanced FIFO slot reservation \textbf{\textit{Algorithm 1}}, they immediately identify their target vehicles and apply the motion control and estimation algorithms. Once they cross the current intersection, the reserved slots are reset to zero, waiting for new assignments while approaching the next intersection. It can be noticed from this plot that, the ego vehicle (i.e., dark-red dashed line) is assigned a new slot of 3 after it enters the next link, even before NPC vehicle 5 and 6 cross the first intersection. This means the proposed slot reservation process is independent at each intersection, which continues running when different vehicles approaching and leaving the intersection.

\begin{figure}
    \centering
    \includegraphics[width=0.8\columnwidth]{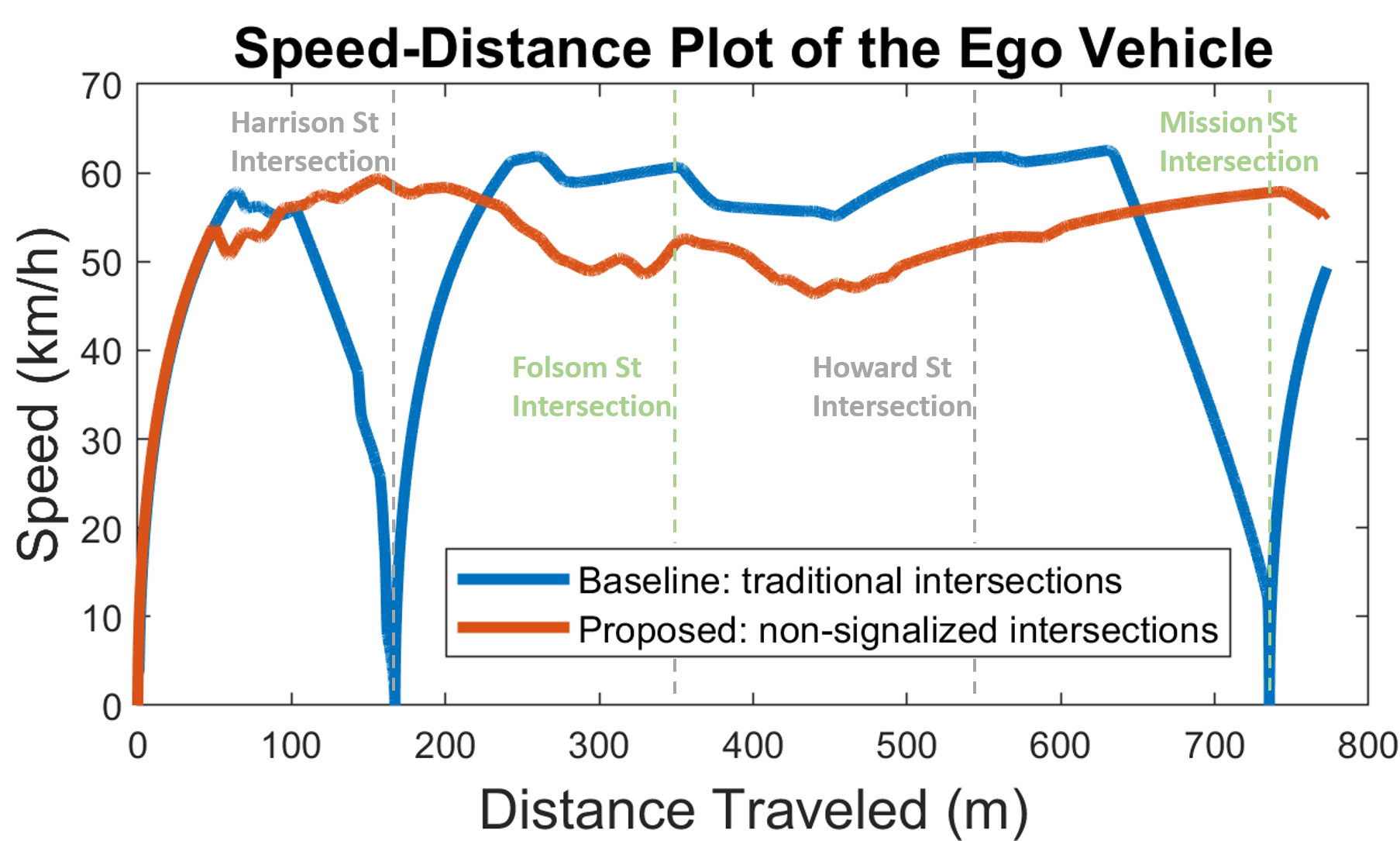}
    \caption{A sample comparison when the ego vehicle travels through the whole corridor 1) with fixed-timing signalized intersections by traditional driving and 2) with non-signalized intersections by \first{the Digital Twin-assisted} cooperative driving}
    \label{DV}
\end{figure}

A sample comparison between the proposed non-signalized intersection scenario and the fixed-timing signalized intersection scenario is shown in Fig. \ref{DV}. This speed-distance plot is created when the same participant drives the ego vehicle through all four intersections. In the baseline scenario, the ego vehicle runs into red lights at the first and the fourth intersections, while directly passes the second and the third intersections during green lights. In the non-signalized intersection scenario, however, the ego vehicle maintains a relatively stable speed while travelling through all four intersections, without any full stop at any intersection. Although a higher maximum speed is reached in the baseline scenario, the excessive speed changes significantly increase the travel time and the energy consumption. 

Based on all trips conducted by the participants in HITL simulation, an average of 20\% reduction in travel time, and an average of 23.7\% reduction in fuel consumption (calculated by the open-source MOVESTAR model \cite{wang2020movestar} and assuming all gasoline vehicles) can be achieved by applying the proposed Digital-Twin assisted cooperative driving system, \first{compared to the baseline signalized scenario without the Digital Twin assistance.}

\begin{figure}%
\centering
\subfigure[]{%
\includegraphics[height=1.35in]{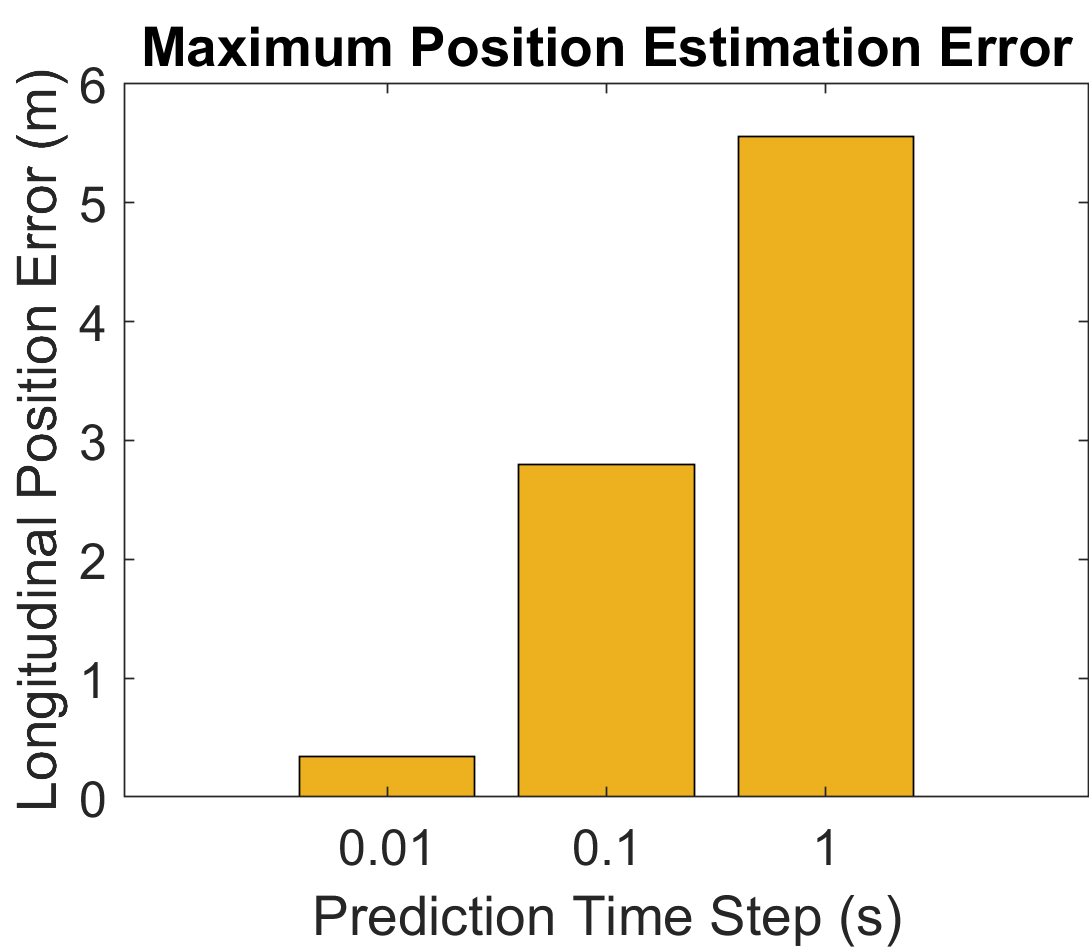}
\label{fig:errormax}
}%
\qquad
\subfigure[]{%
\includegraphics[height=1.35in]{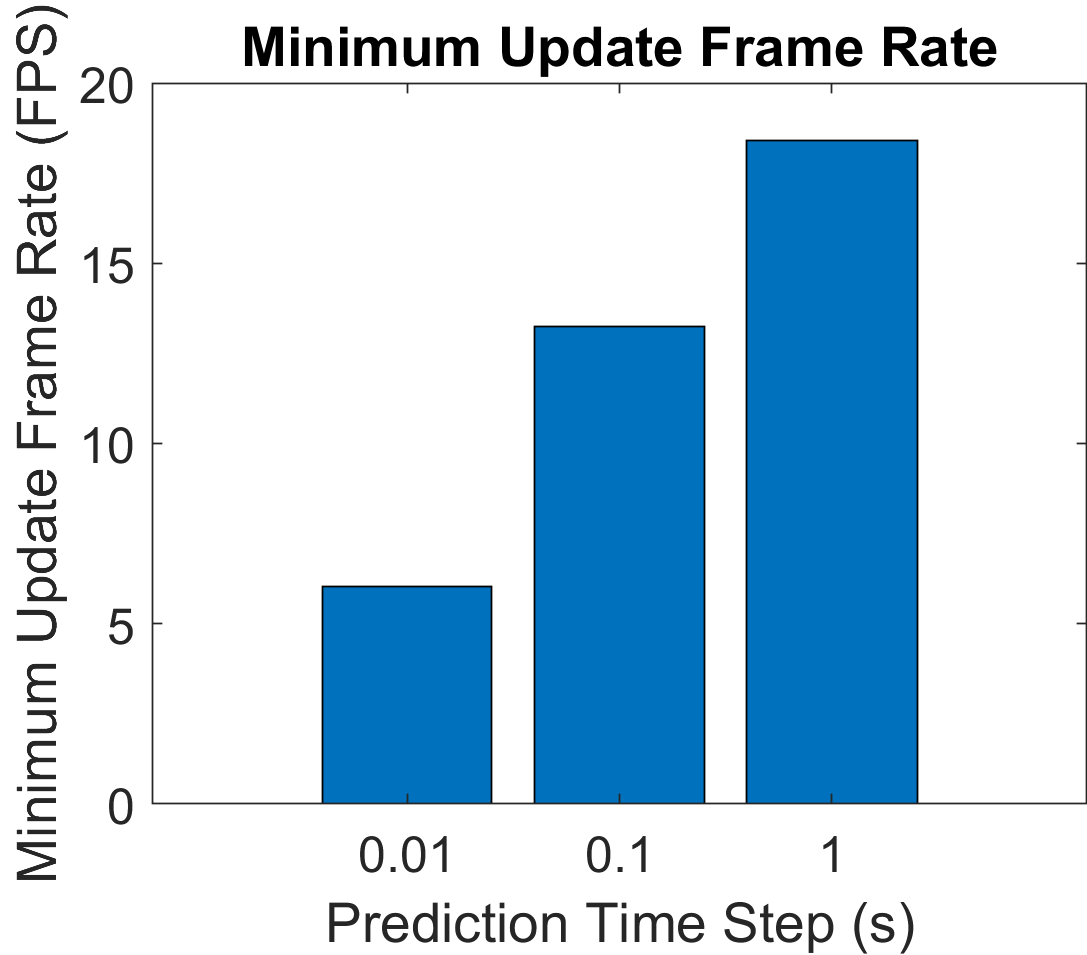}
\label{fig:FPS}
}%
\caption{Simulation results: (a) Maximum error of the ego vehicle's position estimation with different prediction time step values; (b) Minimum update frequency with different prediction time step values}
\end{figure}

\subsection{Sensitivity Analysis on Motion Estimation Algorithm}
Besides HITL simulation, we also conduct sensitivity analysis on our model-based motion estimation algorithm, where we specifically investigate the impacts of different prediction time steps of \textbf{\textit{Algorithm 2}}. \first{``Maximum position estimation error'' is calculated during each trip as the maximum difference between the estimated longitudinal position $\mathbf{\Tilde{D}_n(t)}$ and the ground truth $\mathbf{{D}_n(t)}$. ``Minimum update frame rate'' is calculated during each trip as the minimum frame per second (FPS) value of the simulation.} All connected vehicles \first{in this analysis} are applied with the time-variant communication delay and the hybrid packet loss introduced in Subsection V.A. 

As shown in the simulation results Fig. \ref{fig:errormax} from multiple trips, when the prediction happens more frequently, the maximum position estimation error is decreased. Particularly, when the prediction time step is 0.01 s (i.e., prediction frequency is 100 Hz), the estimation error is always less than 0.2 m based on the given simulation settings of communication delay and packet loss. However, when the prediction time step is 1 s (i.e., prediction frequency is 1 Hz), the estimation error can reach 5.8 m during the major packet loss. This is caused by the difference of information granularity, where a more frequent prediction also means a more frequent information update, which could happen immediately after the packet transmission is resumed from packet loss.

\first{On the other hand}, a more frequent and accurate prediction also leads to a higher computational load. Since this simulation is conducted in Unity game engine, the computational load is measured by the minimum update frame rate of the simulation. As shown in Fig. \ref{fig:FPS}, When the prediction time step is 0.01 s, the frame rate drops to a minimum of 6 FPS, indicating the simulation can be run only six frames per second at that time instant (with existing simulation hardware settings). However, when the prediction time step is 1 s, the frame rate is always higher than 19 FPS. This is caused by the difference of iterations that the proposed motion estimation \textbf{\textit{Algorithm 2}} needs to be called. 

However, since this simulation is conducted on a single computer for multiple vehicles in a centralized fashion, the concern of computational load can be further relieved if cloud computing or distributed computing can be adopted in the future. How to find the trade-off between the prediction accuracy and computational load becomes a critical point of this model-based motion estimation algorithm, \first{which will be further investigated in our following study}.

\section{Conclusion and Future Work} \label{sec:Con}
In this study, the Digital Twin technology has been leveraged to design a cooperative driving system at non-signalized intersections, where connected vehicles cooperate with each other to cross intersections without any full stops. In the digital world of the Digital Twin architecture, the enhanced FIFO slot reservation algorithm has been developed to schedule vehicles' sequences of crossing the intersection, the consensus motion control algorithm has been introduced to generate the referenced motion of vehicles, and the model-based motion estimation algorithm has been proposed to tackle V2X communication issues. In the physical world, a Digital Twin-assisted AR HMI has been designed to provide the guidance to the driver to cooperate with other connected vehicles.     

Agent-based modeling and simulation of the proposed cooperative driving system have been conducted in Unity game engine based on a real-world map in San Francisco. HITL simulation results have proved the benefits of the proposed algorithms with 20\% reduction in travel time and 23.7\% reduction in energy consumption, respectively, when compared with traditional signalized intersections. Sensitivity analysis has also been conducted to study the impacts of the proposed motion estimation algorithm's prediction time step on prediction accuracy and computational load, respectively.

To take this study one step further, legacy vehicles and vulnerable road users (e.g., pedestrians and bicycles) need to be considered in the modeling and visualization process. Particularly, the term ``mixed traffic environment'' does not only refer to an environment mixed with different kinds of vehicles, but also includes vulnerable road users that have the highest priority in the environment. How to build cooperative driving systems that allow connected vehicles to cooperate with legacy vehicles, bicycles and pedestrians at the same time remains an challenging question to be solved.

Additionally, data measured by on-board perception sensors on connected vehicles can also be leveraged in the future to provide additionally sources of motion estimation, which could significantly decreases the motion estimation error of the proposed motion estimation algorithm. \first{And comparisons among our proposed method with other state-of-the-art methods related to intersection control, such as ``Eco-Approach and Departure'' and dynamic/adaptive signal control with connected vehicles, would be helpful to further validate the effectiveness of our study.}

\section*{Acknowledgment}
The authors want to sincerely thank Dr. Bin Cheng and Dr. Haoxin Wang for their guidance on this study, and also thank Kenichi Murata, Akio Orii, Kazuhisa Shitanaka, and Dr. Nejib Ammar for their advice and feedback.

The contents of this study only reflect the views of the authors, who are responsible for the facts and the accuracy of the data presented herein. The contents do not necessarily reflect the official views of Toyota Motor North America.

\ifCLASSOPTIONcaptionsoff
  \newpage
\fi

\bibliographystyle{IEEEtran}
\bibliography{Reference.bib}

\vskip 0pt plus -1fil
\begin{IEEEbiography}
[{\includegraphics[width=1in,height=1.25in,clip,keepaspectratio]{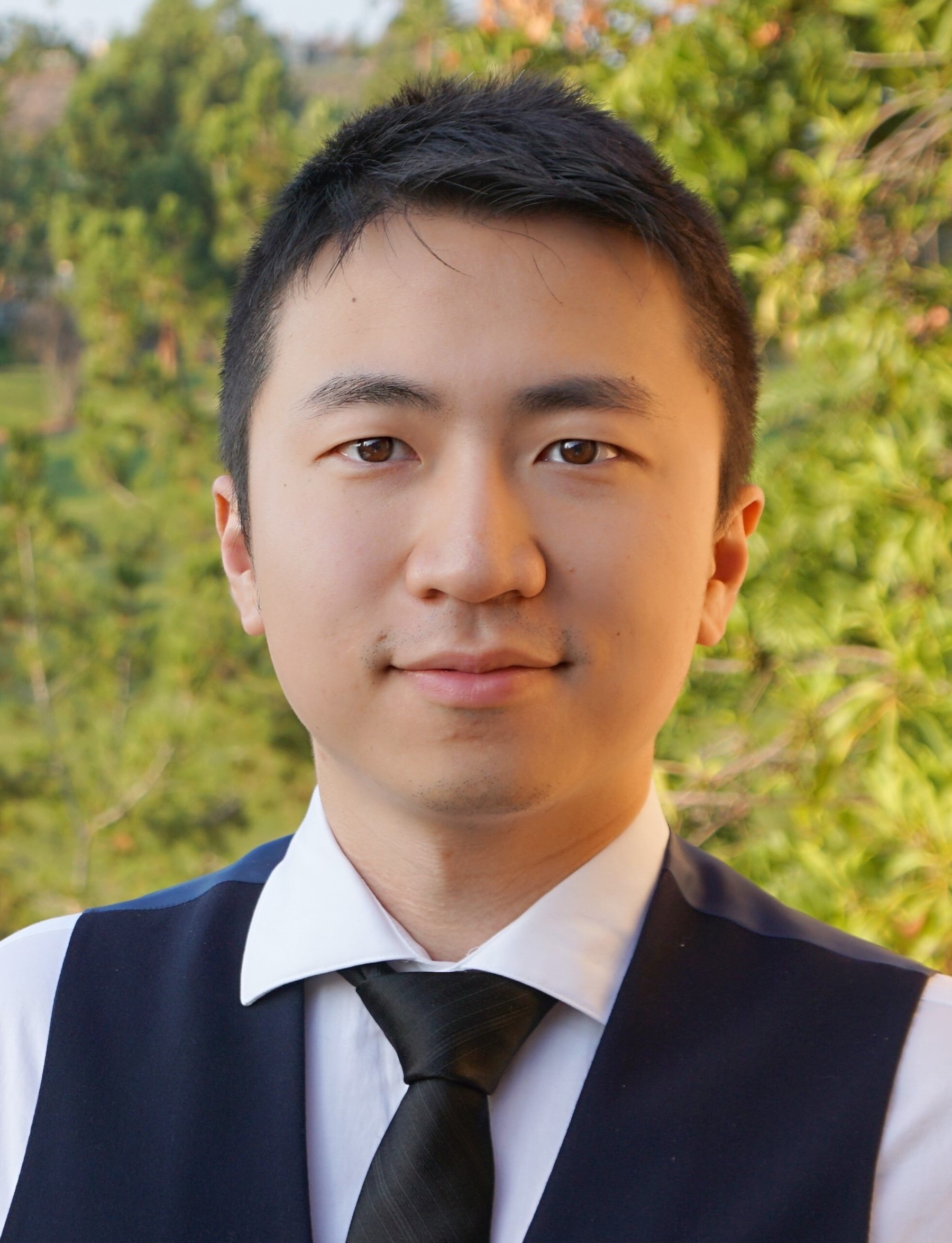}}]
{Ziran Wang}
(S'16-M'19) received the Ph.D. degree from the University of California, Riverside in 2019, and the B.E. degree from Beijing University of Posts and Telecommunications in 2015, respectively. Dr. Wang is currently a Research Scientist at Toyota Motor North America, InfoTech Labs in Silicon Valley, where he conducts research in the ``Digital Twin'' project. His research focuses on intelligent vehicle technology, including cooperative automated driving, driver behavior modeling, and vehicular cyber-physical systems.

Dr. Wang serves as associate editor of SAE International Journal of Connected and Automated Vehicles, founding chair of IEEE Technical Committee on Internet of Things in Intelligent Transportation Systems (IoT in ITS), and member of four other technical committees across IEEE and SAE. Dr. Wang has received the National Center for Sustainable Transportation Dissertation Award from U.S. Department of Transportation, and the Vincent Bendix Automotive Electronics Engineering Award from SAE.
\end{IEEEbiography}

\vskip 0pt plus -1fil
\begin{IEEEbiography}
[{\includegraphics[width=1in,height=1.25in,clip,keepaspectratio]{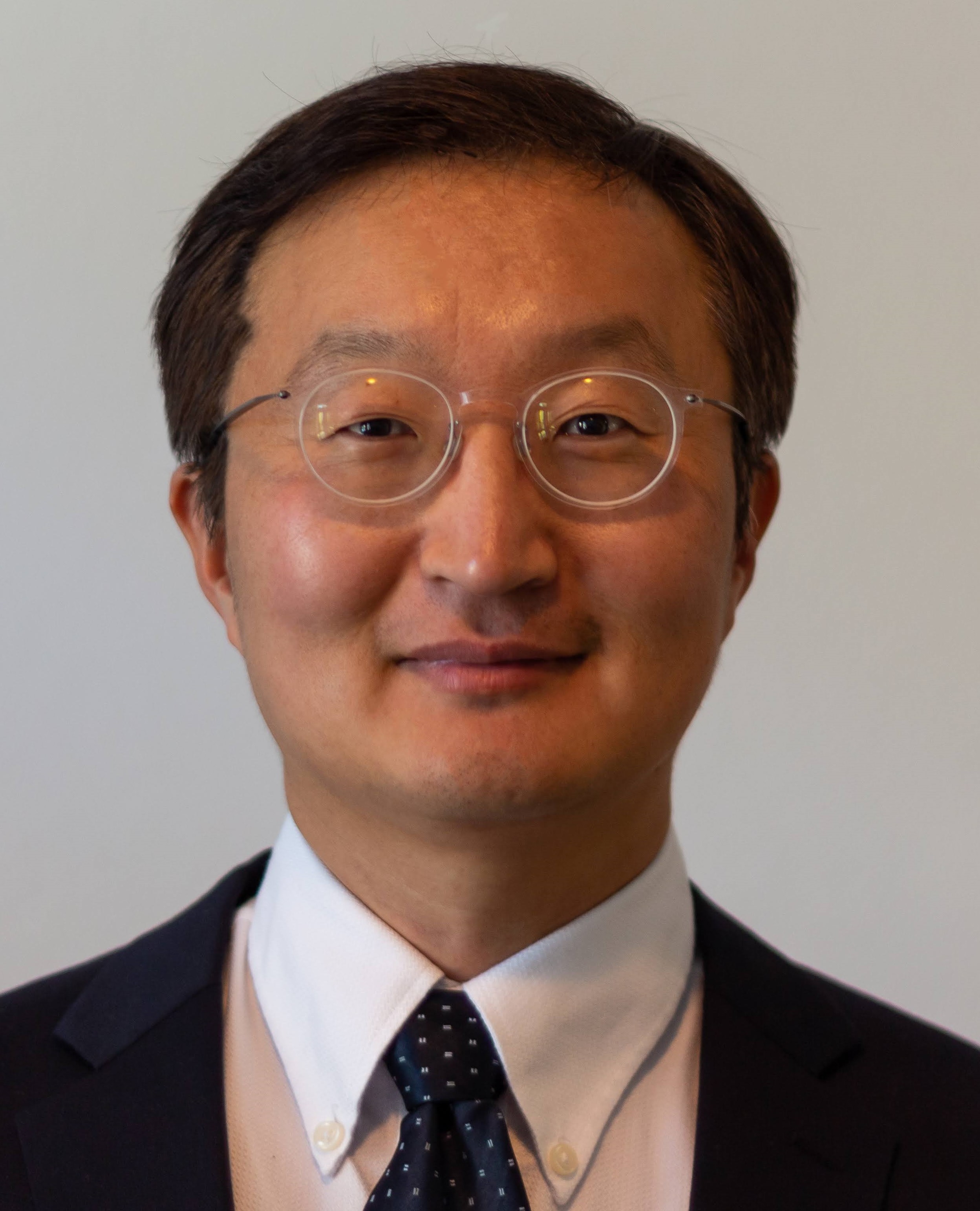}}]
{Kyungtae (KT) Han}
(M'97-SM'15) received the Ph.D. degree in electrical and computer engineering from The University of Texas at Austin in 2006. He is currently a Principal Researcher at Toyota Motor North America, InfoTech Labs. Prior to joining Toyota, Dr. Han was a Research Scientist at Intel Labs, and a Director in Locix Inc. His research interests include cyber-physical systems, connected and automated vehicle technique, and intelligent transportation systems.  
\end{IEEEbiography}

\vskip 0pt plus -1fil
\begin{IEEEbiography}
[{\includegraphics[width=1in,height=1.25in,clip,keepaspectratio]{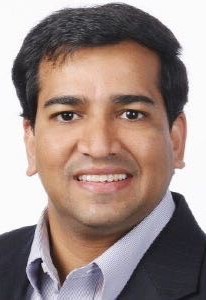}}]
 {Prashant Tiwari}
(M'19) received the Ph.D. degree in mechanical engineering from Rensselaer Polytechnic Institute in 2004, and the MBA degree from University of Chicago in 2016. He is currently a Executive Director at Toyota Motor North America, InfoTech Labs. Dr. Tiwari is highly active in Automotive Edge Computing Consortium (AECC) and SAE. Prior to joining Toyota, Dr. Tiwari held several leadership positions of increasing responsibilities at GE and UTC Aerospace Systems.  
\end{IEEEbiography}

\end{document}